\documentclass[10pt,conference,review]{IEEEtran}
\IEEEoverridecommandlockouts
\usepackage[style=ieee, mincitenames=1, maxcitenames=1, maxbibnames=8]{biblatex}
\bibliography{main}
\usepackage{amsmath,amssymb,amsfonts}
\usepackage{algorithmic}
\usepackage{graphicx}
\usepackage{textcomp}
\usepackage{xcolor}
\usepackage{enumerate}
\usepackage{listings}
\usepackage{booktabs}
\usepackage{multirow}
\usepackage{array}
\usepackage[inline]{enumitem}
\usepackage{adjustbox}
\definecolor{main}{HTML}{5989cf}    
\definecolor{sub}{HTML}{cde4ff}     
\usepackage[many]{tcolorbox}    	
\newtcolorbox{RQanswer}[1][]{%
    colback=sub,
    colframe=black!5,
    notitle,
    sharp corners,
    borderline west={2pt}{0pt}{main!80!black},
    enhanced,
    breakable,
    right=0pt,
    top=0pt,
    bottom=0pt,
    left=2pt
    }
\usepackage{hyperref}
\hypersetup{colorlinks,allcolors=black}

\usepackage{enumitem}
\newlist{questions}{enumerate}{2}
\setlist[questions,1]{label=RQ\arabic*:,ref=RQ\arabic*,leftmargin=*,labelindent=0em}
\setlist[questions,2]{label=(\alph*),ref=\thequestionsi(\alph*)}
\newtheorem{definition}{Definition}

\begin{document}

\title{
How Much Do Code Language Models Remember? 
An Investigation on Data Extraction Attacks before and after Fine-tuning
}

\author{\IEEEauthorblockN{Fabio Salerno}
 \IEEEauthorblockA{\textit{Delft University of Technology}\\
 Delft, The Netherlands \\
 f.salerno@tudelft.nl}
 \and
 \IEEEauthorblockN{Ali Al-Kaswan}
 \IEEEauthorblockA{\textit{Delft University of Technology}\\
 Delft, The Netherlands \\
 a.al-kaswan@tudelft.nl}
 \and
 \IEEEauthorblockN{Maliheh Izadi}
 \IEEEauthorblockA{\textit{Delft University of Technology}\\
 Delft, The Netherlands \\
 m.izadi@tudelft.nl}
 }

\maketitle

\begin{abstract}
Code language models, 
while widely popular, 
are often trained on unsanitized source code 
gathered from across the Internet.
Previous work revealed that 
\textit{pre-trained} models can
remember the content of their training data 
and regurgitate them 
through data extraction attacks.
Due to the large size of current models, 
only a few entities have the resources for pre-training such models. 
However, fine-tuning requires fewer resources and 
is increasingly used by both small and large entities 
for its effectiveness on specialized data.
Such small curated data for fine-tuning 
might contain sensitive information or proprietary assets.
In this study, we attack 
\textit{both} pre-trained and fine-tuned code language models 
to investigate the extent of data extractability. 
We first develop a custom benchmark to assess the vulnerability of both 
pre-training and fine-tuning samples to extraction attacks. 
Our findings reveal that 54.9\% of extractable pre-training data 
could be retrieved from StarCoder2-15B, 
whereas this number decreased to 23.5\% after fine-tuning.
This indicates that fine-tuning reduces the extractability of \textit{pre-training} data.
However, compared to larger models, 
\textit{fine-tuning smaller} models 
increases their vulnerability to data extraction attacks 
on \textit{fine-tuning} data.
Given the potential sensitivity of fine-tuning data, 
this can lead to more severe consequences.
Lastly, we also manually analyzed 
2000 extractable samples before and after fine-tuning. 
We also found that data carriers 
and licensing information 
are the most likely data categories to 
be memorized from pre-trained and fine-tuned models, 
while the latter is the most likely to be forgotten after fine-tuning. 
\end{abstract}


\section{Introduction}
Large Language Models (LLMs) have become increasingly popular for their ability to generate human-like text and their potential applications across various fields, including Software Engineering~\cite{hou2023large, lu2021codexglue,izadi2024language}. 
Despite their immense popularity, LLMs for code are commonly trained on \textit{unsanitized} datasets of source code scraped from the Internet. Previous work has revealed that the content of these datasets can be memorized and extracted by attackers through various data extraction techniques~\cite{carlini2021extracting, biderman2023emergent,alkaswan2024traces,ishihara2023training, yang2024unveiling}.

While LLM pre-training typically occurs infrequently due to their enormous size, fine-tuning is becoming the dominant approach for end-users to customize models. Fine-tuning often involves small curated datasets~\cite{bogomolov2022assessing}, which may contain proprietary or sensitive  information~\cite{alkaswan2023abuse, alkaswan2024traces}. This raises concerns about the potential for sensitive data within these datasets to be inadvertently extracted from the fine-tuned models~\cite{mireshghallah2022memorization, zeng2023exploring, sun2023does}.
To our knowledge, previous studies have investigated data memorization and extraction attacks on \textit{pre-trained} LLMs for code~\cite{alkaswan2024traces, zeng2023exploring}, but there has been no empirical investigation into the extractability of \textit{fine-tuned} code models. In this work, we explore the extent to which fine-tuned LLMs for code leak data from their pre-training and fine-tuning corpora.

As there is no comprehensive framework or approach for measuring memorization, we start by defining a data extraction security game grounded in the theory behind membership inference attacks and k-extractability~\cite{carlini2021extracting, carlini2022membership}. Using this security game, we develop a framework to quantify memorization in LLMs, using extractability as a proxy. While memorization of training data can manifest as non-exact duplication~\cite{hartmann2023sokmemorizationgeneralpurposelarge}, measuring the rate of data extraction provides a lower bound on data leakage in LLMs.

We first develop a custom benchmark to assess the vulnerability of pre-training and fine-tuning samples to extraction attacks and identify vulnerable samples. We then test the extractability of pre-training and fine-tuning data on the Starcoder2 model family~\cite{lozhkov2024starcoder2stackv2}. Finally, two authors individually and manually categorize the extractable samples to identify specific data categories that are more exposed to memorization.

Our key results: \textit{Fine-tuned LLMs for code tend to ``forget'' samples that were extractable from the pre-trained models. Additionally, smaller models are more vulnerable to fine-tuning data extraction attacks compared to larger models. Data carriers and license information are the most likely to be memorized, while the latter is the most likely to be forgotten after fine-tuning, leaving data carriers still exposed to extraction.}

To summarize, the main contributions of our study are:
\begin{itemize}
    \item An empirical assessment of pre-training data memorization in fine-tuned code models which reveals that fine-tuning reduces the extractability of certain pre-training data from the pre-trained models.
    \item An empirical assessment of fine-tuning data memorization in code models which shows that (1) code models memorize fine-tuning data; (2) smaller models are more affected by fine-tuning memorization; (3) increasing the duplication rate of samples from the training corpus significantly drives data leakage; (4) similar to pre-training memorization, the input prefix length has a considerable impact on fine-tuning intractability. 
    \item We release our code, evaluation scripts, and data used in this study to enable replication of our results ~\cite{rep_package}. \footnote{GitHub repo: \url{https://github.com/AISE-TUDelft/LLM4Code-memtune}}
\end{itemize}

\section{Background and Related Work}
\paragraph{Memorization}
Memorization refers to the ability of Language Models to recall details from the training data~\cite{carlini2021extracting, carlini2022membership, carlini2022privacy, alkaswan2023abuse, yang2024unveiling, hu2022membership}. Recent studies have highlighted this risk of regurgitating training data, some of which might contain sensitive information~\cite{alkaswan2023abuse, carlini2021extracting}. 

\paragraph{Membership Inference Attacks}
Memorization enables membership inference attacks~\cite{alkaswan2023Targeted, carlini2022membership, carlini2022privacy}. With a membership inference attack, an adversary aims to predict if a sample was part of the training data of a machine learning model. The adversary is only given access to the model, and no access to the training data. Membership inference attacks are a primitive of more elaborate and dangerous attacks against machine learning models, such as data extraction attacks.
\citeauthor{shokri2017membership} proposed the first membership inference attack in which the authors targeted classification models~\cite{shokri2017membership}. Since then the field has expanded and proposed attacks target generative models, including LLMs~\cite{carlini2022membership, mireshghallah2022quantifying, oh2023membershipKo}.
In the code domain, an additional risk is raised by the presence of copyleft and other non-permissively licensed codes in training datasets~\cite{alkaswan2023abuse}. \citeauthor{katzy2024exploratory} found that datasets used to train code LLMs contain non-permissively licensed code, even those that claim otherwise~\cite{katzy2024exploratory}.    

\paragraph{Untargeted Data Extraction Attacks}
~\citeauthor{yang2024unveiling} studied memorization in \textit{pre-trained} code LLMs using an untargeted attack against CodeParrot \cite{yang2024unveiling}. In untargeted attacks, the model is prompted with an empty string or a random token. The completion of the model is then checked for memorized sequences. The memorized sequence can be from any part of the training corpus~\cite{carlini2021extracting, oh2023membershipKo, carlini2022privacy}. This attack requires knowledge of the complete training data, as any random sequence can be emitted, as there is no control of the target~\cite{alkaswan2023Targeted, henderson2023foundation,huang2022large, mireshghallah2022quantifying}. 

\paragraph{Targeted Data Extraction Attacks}
\citeauthor{carlini2021extracting} first proposed a targeted data extraction attack and demonstrated that LMs tend to memorize and regenerate segments of training data. \citeauthor{kandpal2022deduplicating} and \citeauthor{lee2022deduplicating} revealed that duplicated training data is more vulnerable to memorization, and de-duplication can effectively reduce memorization~\cite{kandpal2022deduplicating, lee2022deduplicating}. \citeauthor{carlini2022quantifying} further quantified memorization effects, exhibiting that memorization grows with model scale, data duplication, and prompt input length~\cite{carlini2022quantifying}. 
\citeauthor{alkaswan2024traces} also measured memorization in \textit{pre-trained} code LLMs but with a targeted attack.
In a targeted attack, the model is supplied with partial information to complete. 
By sampling common code from Python files and identifying potentially extractable samples, the authors constructed a benchmark dataset~\cite{alkaswan2024traces}.

\paragraph{Fine-tuning}
Many code LLMs are deployed using the pretrain-finetune paradigm. Under this paradigm, an LLM is first pre-trained with a large unlabeled dataset in an unsupervised manner. Then, the LLM is fine-tuned with a smaller task-specific dataset~\cite{devlin2019bert, wang-etal-2021-codet5,van2023enriching}. 
Use cases for fine-tuning include adapting a model to a specific project~\cite{ahmed2022few}, organization, or a different field~\cite{alkaswan2023extending}. Additionally, LLMs can be fine-tuned with instruction data~\cite{zhang2023instruction}.

\paragraph{Memorization after Fine-tuning}

Research on the impacts of fine-tuning on memorization is restricted and it has largely focused on text-based models.
\citeauthor{mireshghallah2022memorization}~\cite{mireshghallah2022memorization} experimented with different fine-tuning methods on GPT-2 to investigate their impact on memorization. 
\citeauthor{zeng2023exploring} found that the fine-tuning task also significantly influences the memorization rate~\cite{zeng2023exploring}. 

\citeauthor{sun2023does}~\cite{sun2023does} investigate the fine-tuning API offered by GPT-3~\cite{brown2020language}. A fine-tuned GPT-3 text completion model can leak 
Personal Identifiable Information (PII) from its pre-training data~\cite{sun2023does}.

Research on code-based LLMs is even more limited. \citeauthor{pappu2024measuring} investigated memorization in Reinforcement Learning with Human Feedback (RLHF) for code completion models~\cite{pappu2024measuring}. Memorization of pre-training data can persist through RLHF, but the memorization rate of the RLHF data is much lower~\cite{pappu2024measuring}. To the best of our knowledge, no prior studies have examined how fine-tuning affects memorization in code LLMs.

\paragraph{Our contribution}
In our study, we examine fine-tuning and its impact on memorization in code models. We build on previous findings by exploring \textit{multiple factors} well studied for pre-trained code models but not yet for fine-tuned ones, such as input prefix lengths, duplication rates, and model sizes. We also \textit{manually} investigate a set of extracted data to understand the type of memorized information. While previous research primarily focused on the memorization of Python code, our work uniquely explores the memorization of \textit{Java} code, which has not been studied before.

\section{Problem Definition}
Previous studies have shown that LLMs trained on code tend to memorize parts of their training data~\cite{alkaswan2024traces, yang2024unveiling, pappu2024measuring}. Fine-tuning these models often uses small, curated datasets, which are often proprietary and valuable. 
This poses additional risks of sensitive information being unintentionally exposed through the fine-tuned models~\cite{mireshghallah2022memorization, sun2023does, zeng2023exploring}.
In this work, using data extractability techniques, 
we investigate how much code LLMs 
memorize the data they have seen before and after \textit{fine-tuning}.
 
We examine this phenomenon for the automatic code completion task, 
which has emerged as a highly prevalent and valuable application of LLMs~\cite{hou2023large}.
Code completion also represents a setting where leakage of user data can raise legal, commercial, and privacy concerns~\cite{alkaswan2023abuse}.
To this end,
we fine-tune the StarCoder2 model~\cite{lozhkov2024starcoder2stackv2} and its various available parameter sizes into a downstream task; line completion in Java. 
We focus on Java for several reasons. 
Java is widely used in enterprise environments, making it a relevant and practical choice. 
Furthermore, prior studies on code memorization have predominantly focused on Python and suggested that more structured languages might show different patterns of extractability~\cite{alkaswan2024traces, yang2024unveiling, pappu2024measuring}.  Following their recommendations, we chose to investigate Java as a new language.

\begin{figure}
    \centering
    \includegraphics[width=\linewidth]{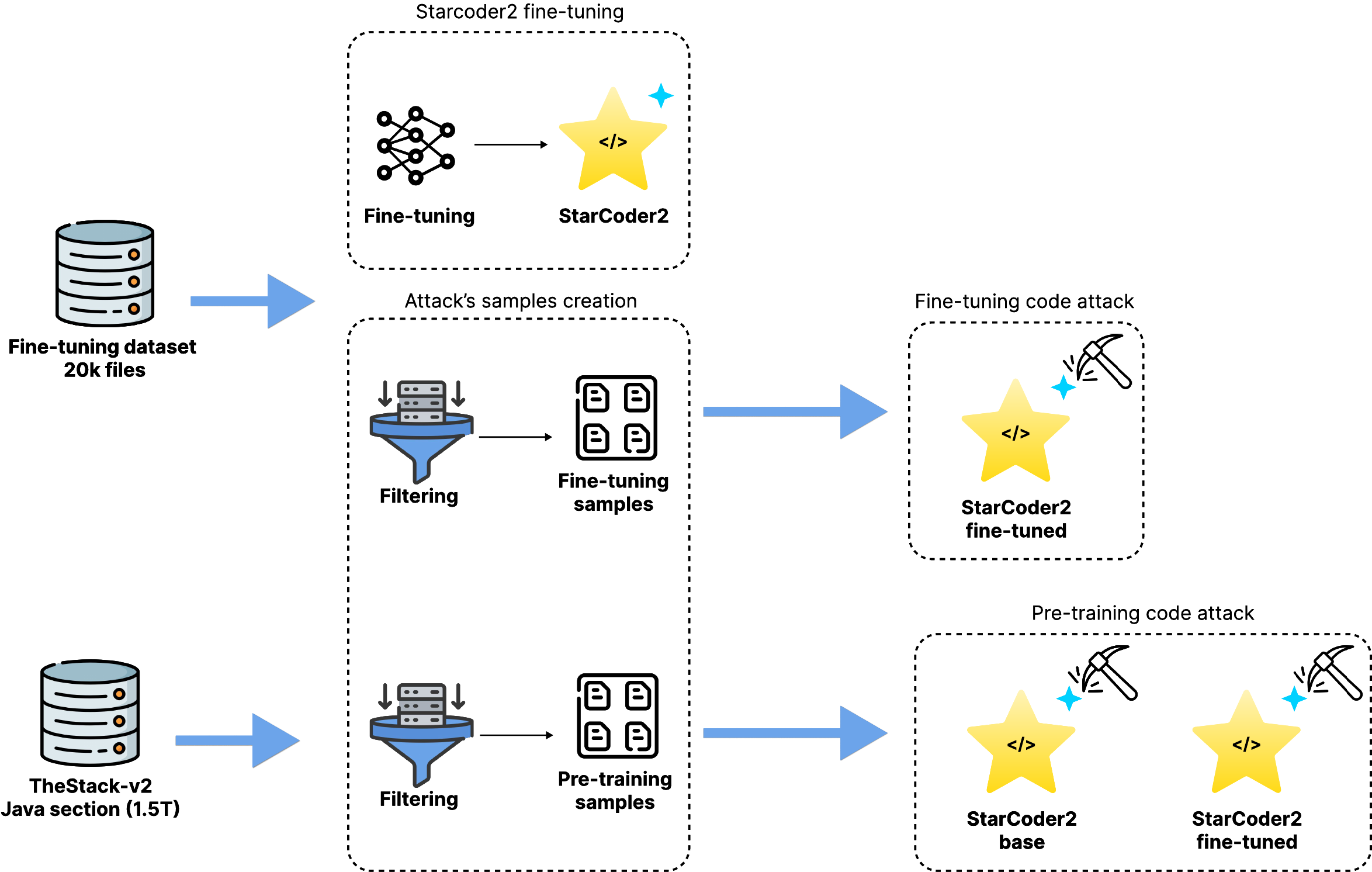}
    \caption{Overview of the experiments' configuration.}
    \label{fig:paper-sum}
\end{figure}

\autoref{fig:paper-sum} provides an overview of the configuration of the experiments. We adopt a setup introduced by \citeauthor{carlini2022privacy}~\cite{carlini2022privacy} to conduct targeted data extraction attacks. Targeted attacks are more security-critical as they allow for targeting specific information from the training corpus. We perform two types of targeted attacks: a \textit{pre-training code attack} to evaluate the extractability of pre-training data, and a \textit{fine-tuning code attack} to quantify the vulnerability of recent data encountered by the model.

\subsection{StarCoder2 and TheStack v2}
For this study, we utilized StarCoder2~\cite{lozhkov2024starcoder2stackv2}. StarCoder2 is a transparently trained open code LLM and comes in three sizes: 3B, 7B, and 15B parameters. 
All models use Grouped Query Attention, a context window of $16384$ tokens with sliding window attention of $4096$ tokens, and were trained using the Fill-in-the-Middle objective~\cite{lozhkov2024starcoder2stackv2}. 

The Stack v2 is the largest open code dataset suitable for LLM pre-training; it builds on Software Heritage's extensive source code archive, which covers over 600 programming languages. In addition to code repositories, The Stack v2 includes curated data from high-quality open sources such as GitHub issues, pull requests, Kaggle and Jupyter notebooks, code documentation, and other natural language datasets related to mathematics, coding, and reasoning. The complete dataset comprises approximately 67.5 TB and around 900 billion tokens.\footnote{TheStack v2: \url{https://huggingface.co/datasets/bigcode/the-stack-v2}}
We chose StarCoder2 as it is one of the most recent open-source code models. Starcoder2 and its derivatives are the best-performing models with both open weights and open data on the EvalPlus Leaderboard~\cite{evalplus}.\footnote{EvalPlus Leaderboard: \url{https://evalplus.github.io/leaderboard.html}} Its open-data nature provides us with complete access to the training corpus, which is essential for conducting targeted attacks and assessing the extractability of pre-training data~\cite{yang2024unveiling, alkaswan2024traces}.

\subsection{Research Questions}
\begin{questions}
    \item\textbf{To what extent are pre-training data extractable from pre-trained models?} We conduct a data extraction attack on the base version of StarCoder2 to evaluate the extractability of its training data. We examine two factors that might influence extractability: model size and input prefix length ~\cite{alkaswan2024traces, carlini2021extracting}.

    \item \textbf{To what extent are pre-training data extractable from fine-tuned models?} After fine-tuning the model on the Java line completion task, we repeat the pre-training code attack to compare the extractability before and after fine-tuning.
    
    \item \textbf{To what extent are fine-tuning data extractable from fine-tuned models?} We conduct a data extraction attack on the fine-tuned StarCoder2 to evaluate the extractability of recent fine-tuning data. Factors analyzed include model size, training epochs, duplication rate of samples, and input prefix length variations ~\cite{alkaswan2024traces, carlini2021extracting}.

    \item \textbf{Does the type of code data impact extractability?} Finally, we investigate whether certain data categories are more or less extractable after fine-tuning. We analyze samples from both fine-tuning and pre-training data extraction attacks, constructing a classification to evaluate their extractability.

\end{questions}

\section{Fine-tuning}
\subsection{Dataset}
The fine-tuning dataset consists of 20k \texttt{.java} files randomly sampled from the Java subset of The Heap\cite{katzy2025heapcontaminationfreemultilingualcode}.\footnote{The Heap Dataset: \url{https://huggingface.co/datasets/AISE-TUDelft/the-heap}} This dataset was created by scraping 10.5k public repositories on GitHub. The Heap consists of repositories released under strong copyleft licenses, as these licenses are excluded from the StarCoder2 training corpus~\cite{lozhkov2024starcoder2stackv2, katzy2024exploratory, katzy2025heapcontaminationfreemultilingualcode}.
The scraped files are near-deduplicated against The Stack to remove any possibility of training-data contamination~\cite{katzy2024exploratory, katzy2025heapcontaminationfreemultilingualcode}.

We further apply several filters to the dataset in-line with the StarCoder2 pre-processing pipeline~\cite{lozhkov2024starcoder2stackv2}. We remove files that are too long, have too few alphabetic characters, and those consisting of encoded or autogenerated data~\cite{lozhkov2024starcoder2stackv2}.

\subsection{Training}
The fine-tuning of each model was conducted using the resources provided by the Delft High-Performance Compute Cluster~\cite{DHPC2024}.  
We utilize $32$ CPU cores with $32$GB of RAM and a set of Nvidia A100 GPUs, each with $80$GB of memory. The experiments are performed on various sizes of the StarCoder2 model, with the specific number of GPUs allocated based on the model's parameter count to fit within memory constraints. Specifically, for StarCoder2-3B, 7B, and 15B, we employ two, four, and six GPUs, respectively.
This results in approximate training times of $25$, $55$, and $110$ hours, respectively. The GPUs are running Nvidia driver version $555.42.02$ and CUDA $12.5$. The training utilizes Transformer version $4.41.1$ on Torch $2.3.0$+cu$121$.

To ensure consistent comparison across the different sizes of StarCoder2, we maintain the same training setup for each model size. 
The effective training batch size (including gradient accumulation) is $24$, $24$, and $25$, respectively. 
The context window is set to $1024$ tokens. We use a learning rate of $3e-5$ and utilize a linear scheduler and the Adafactor optimizer~\cite{shazeer2018adafactor} to reduce GPU memory requirement and to make reproducibility easier.
Each model was trained for three epochs, with checkpoints saved at the end of each epoch. 
To avoid overfitting, we monitored the validation loss after each fine-tuning epoch to ensure it did not increase. This approach confirms that the models are not overfitting.
\footnote{The loss graphs are available in the replication package \cite{rep_package}: training/train-stats}

Detailed training configurations and results are available in the replication package~\cite{rep_package}. 

\section{Data Extraction Attacks}
We first introduce a precise definition of memorization proposed by \citeauthor{carlini2022quantifying}~\cite{carlini2022quantifying}.
\begin{definition}[memorization]
\label{def:mem}
    A string $s$ is \textit{extractable with k tokens of context} from a model $f$ if there exists a ($k$-length) string $p$, such that the concatenation $[p||s]$ is contained in the training data for $f$, and $f$ produces $s$ when prompted with $p$ using greedy decoding.
\end{definition}

This is defined as verbatim memorization. \citeauthor{ippolito-etal-2023-preventing}~\cite{ippolito-etal-2023-preventing} proposed a relaxed definition of memorization using the Bilingual Evaluation Understudy (BLEU) score~\cite{ippolito-etal-2023-preventing}. To measure memorization, we first formally define the data extraction game and then describe the approach used to obtain the pre-train and fine-tuning code samples for performing the data extraction attacks.

\subsection{Data Extraction Security Game}
In our experiments, we consider the models as black-box systems. We define a security game inspired by the membership inference attack security game and the notion of k-extractability ~\cite{carlini2021extracting, carlini2022quantifying, alkaswan2024traces}. 

\begin{definition}[Data extraction security game]
\label{def:extgame}
Given a challenger $\mathcal{C}$, an adversary $\mathcal{A}$, a data distribution $\mathbb{D}$ and a model $f$ the game is defined as follows:
\begin{enumerate}
    \item The challenger samples a training dataset $D \gets \mathbb{D}$ and trains a model $f_\theta \gets \mathcal{T}(D)$ on the dataset $D$.
    \item $\mathcal{C}$ samples a sample \(D_n = (p, s)\) where \(D_n \in D\). The prefix \(p\) is provided to the adversary $\mathcal{A}$.
    \item $\mathcal{A}$ is allowed query access to the model \(f_\theta\) and may perform any other polynomial-time operations
    \item $\mathcal{A}$ outputs his prediction sequence \(\hat{s}\)
    \item If \(\hat{s} = s\), $\mathcal{A}$ wins, otherwise $\mathcal{C}$ wins
\end{enumerate}
\end{definition}
In other words, given a prefix $p$ (1), the adversary is challenged to extract the correct suffix from the model's training data. The adversary can query the language model $f_\theta$(2) but cannot inspect the weights or hidden states (e.g., attention vectors). The adversary then predicts the suffix $s$ (3) and wins if it matches the actual suffix in the training data~\cite{carlini2021extracting, carlini2022quantifying, alkaswan2024traces}. 

Some factors may affect the difficulty of the challenge:
\begin{enumerate}
    \item The selection of the sample $D_n \in D$. As noticed by previous works, not all training samples are as hard to extract as others. In particular, highly duplicated or outlier samples are more vulnerable to attacks ~\cite{kandpal2022deduplicating, lee2022deduplicating, carlini2022quantifying}.

    \item The size of the model $f_\theta$. Some models are more likely to memorize samples than others, i.e., larger models have been observed to memorize more pre-training data than smaller models ~\cite{carlini2022quantifying, alkaswan2024traces, yang2024unveiling}. 

    \item The length of the prefix $p$. It has been found that longer prefixes induce more memorization. ~\cite{carlini2022quantifying, yang2024unveiling}.

    \item The victory condition $\hat{s} = s$. Using the exact match might exclude samples that are entirely extractable but have meaningless differences e.g., a space character.
    So, a fuzzy match can also be considered ~\cite{ippolito-etal-2023-preventing}.
\end{enumerate}
In this work, we exploit the security game and the factors (1,3,4) described above to construct a set of extractable samples to evaluate different model sizes (2). We also measure fuzzy match scores (4) and compare them with the exact match rate. 

\subsection{Data Extraction Dataset}
To measure memorization in LLMs for code, we first need to construct a dataset analogous to the one used in the SaTML23 Language Model Data Extraction Challenge.~\footnote{Language Models Training Data Extraction Challenge: \url{https://github.com/google-research/lm-extraction-benchmark}}
We can divide the extractions into 
\textit{pre-training code attack} (extracting pre-training data from the model) 
and \textit{fine-tuning code attack} (extracting fine-tuning data from the model). 
In our experiments, we use different sizes of StarCoder2 to improve generalizability.

We collect the pre-training samples for the attack from the Java subset of TheStack-v2 through the following procedure. The candidate samples for the attack are obtained by randomly selecting 300 token spans from anywhere in the file taking into account the context window used to train StarCoder2~\cite{lozhkov2024starcoder2stackv2}. 
All the files with less than 300 tokens were discarded. 
We considered only candidates with more than three duplicates inside the training corpus to be consistent among the RQs given the space limit. 
Additionally, to avoid overlaps we exclude the candidate samples also included in the fine-tuning set. 
We randomly select 1K samples from the filtered candidate set to perform our evaluation. 
Finally, we split the 300 tokens sequences into a suffix of 50 tokens and a prefix of 250 tokens. 
In our experimental setup, we will truncate the prefix to 100, 150, and 200 tokens to test extractability when feeding the inputs with different prefix lengths to the models. As in \citeauthor{alkaswan2024traces} ~\cite{alkaswan2024traces}, our pre-train attack dataset construction procedure differs from the one used by ~\citeauthor{carlini2022quantifying}~\cite{carlini2022quantifying} in one aspect. 
Our dataset does not guarantee that for every $D_n = (s,p)$ 
there does not exist a $(s,p') \in D$ where $p' \neq p$. 
This is because the computational cost of identifying all unique samples $D_n = (s,p)$ 
is extremely large for a dataset of this size~\cite{alkaswan2024traces}.

The approach for collecting fine-tuning samples for the attack presents variations from the previously described procedure. Given the smaller scale of the fine-tuning set compared to the pre-training set, we implemented certain modifications that enabled further experimentation. 
We apply the candidate samples for the attack by applying a 300 tokens sliding window to each file of the fine-tuning set while considering the context window used in the fine-tuning setup. 
The size of the fine-tuning set allows us to know and control the number of times a 300-token sample is in the fine-tuning set; this enables us to control the duplication rate. 
We divided candidate samples into subsets based on their duplication rate: 
unique samples ($d=1$), single duplicates ($d=2$), double duplicates ($d=3$), 
and samples with more than three duplicates ($d>3$). 
This choice is guided by inspecting the duplication rate sample distribution 
on the fine-tuning set. 

For each duplication rate subset, we randomly select 1K samples from the filtered candidate set to perform our evaluation. Finally, we split the 300 token sequences into a suffix of 50 tokens and a prefix of 250 tokens with the same approach we used for the pre-training attack dataset. 
The smaller scale of the fine-tuning set compared to the pre-training set allowed us 
to guarantee that for every $D_n = (s,p)$, 
there does not exist a \((s,p \text{’}) \in D\) where \(p \neq p\text{’}\) 
as in the attack dataset construction procedure by \citeauthor{carlini2022quantifying}~\cite{carlini2022quantifying}.

\subsection{Performing the Attacks}
We prompt the model under investigation with the prefix. 
We utilize the standard generation pipeline and the default generation configuration of StarCoder2, as defined in the model configuration. 
The models are prompted in a one-shot fashion with greedy decoding. 
The generated suffix is then compared with the true suffix, which serves as the ground truth. We evaluate the performance using two types of metrics. 
First, we measure the Exact Match (EM) score. Additionally, we also assess the match by fuzzier metrics such as BLEU~\cite{post-2018-call}, METEOR~\cite{banarjee2005}, and ROUGE-L~\cite{lin-2004-rouge}. 

\section{Analysis through Categorization}
To better understand the type of extracted data we conduct an exploratory study by categorizing both the pre-trained and fine-tuned attack datasets.
For consistency with the framework in the literature, we decided to keep and use the code categories defined by ~\citeauthor{alkaswan2024traces} ~\cite{alkaswan2024traces}.
The different categories are outlined in Table~\ref{tab:cats}.
For fine-tuning, we selected one of the four 1k code attack datasets, 
specifically choosing the sample set with a duplication rate of 3. 
This choice represents a balance between duplication rate and category representativeness. 

To simplify the classification process, 
samples with dual purposes (e.g., samples that start with license information and finish with code.) were assigned to their predominant category~\cite{alkaswan2024traces}. 

Two authors independently conducted the classification and subsequently compared their results.
Before labeling, both reviewers jointly examine $20$ samples to familiarize themselves with the dataset. 
The labeling process shows strong agreement between annotators, with Cohen’s Kappa scores of $0.87$ for the pre-training code attack dataset and $0.90$ for the fine-tuning code attack dataset.
Analysis of the confusion matrix reveals that discrepancies primarily arise from differences between test categories and code categories, as well as from the decisions made when categorizing samples with dual purposes. We discard the samples that the reviewers do not agree on, and we are left with $899$ and $917$ samples for the pre-training and fine-tuning datasets, respectively.

\begin{table}
    \centering
    \caption{Categories of data extraction datasets.}
    \resizebox{7cm}{!}{%
    \begin{tabular}{ll|ll}
    \noalign{\smallskip}\toprule
    Category        & Purpose               & Pre-training & Fine-tuning     \\
    \cmidrule{1-4}
    Code            & Code Logic            & 388 & 223     \\
    Dicts           & Dictionaries or other data carriers & 164 & 117        \\
    Docs            & Documentation         & 173 & 291         \\
    License         & License information   & 124 & 284        \\
    Testing         & Test Code             & 50 & 2        \\
    \cmidrule{1-4}
    Total           &                       & 899           & 917         \\
    \noalign{\smallskip}\bottomrule
    \end{tabular}}
    \label{tab:cats}
\end{table}

\section{Results}
\subsection{Extracting Pre-training Data from Pre-trained Models}

\begin{figure}
    \centering
    \includegraphics[width=\linewidth]{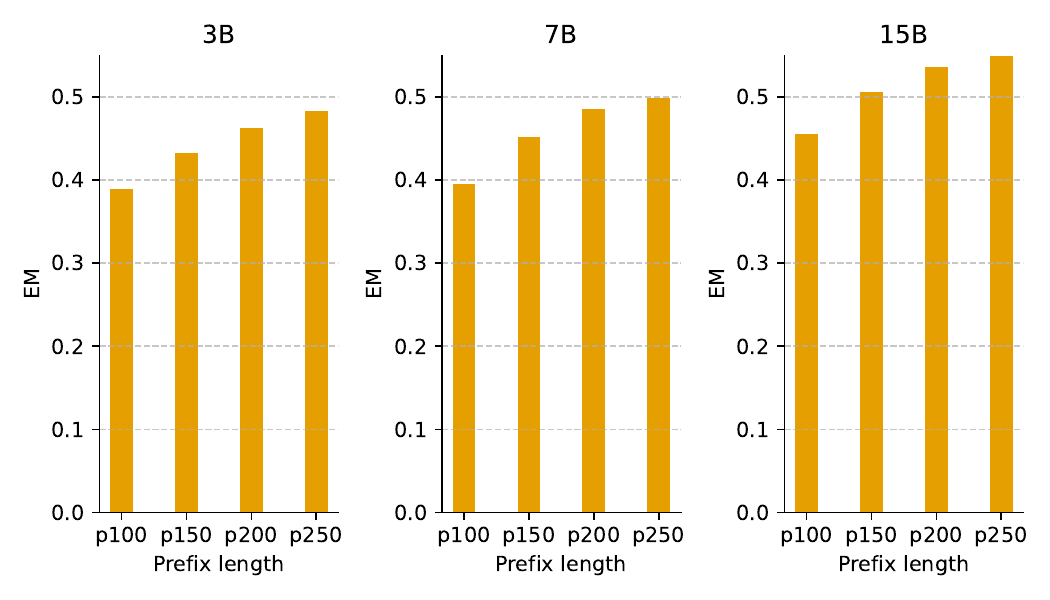}
    \caption{EM rate when performing pre-training code attack on pre-trained models.}
    \label{fig:forg_msize}
\end{figure}

The results of the pre-training code attack on the pre-trained models are shown in Table~\ref{tab:attack-pre-forg-plen}.
We observe that increasing the length of the input prefix in attacks enhances the extractability of pre-training data from pre-trained models.
We are able to extract 45.5\% of the samples from the StarCoder2-15B model using a 100-token prefix.
This extraction rate increases to 50.6\% with a 150-token prefix, and further to 54.9\% with a 250-token prefix. 
This pattern is also supported by analyzing the results of fuzzy metrics.
The suffixes generated by the StarCoder2-15B model reached a BLEU score of 65.2\% with a 100-token prefix, which improved to 72.8\% with a 250-token prefix.
As shown in Figure~\ref{fig:forg_msize}, the increase in extractability is notably more pronounced when extending the prefix length from 100 to 150 tokens compared to the increase from 200 to 250 tokens.
This suggests a nonlinear growth trend concerning prefix length.

\begin{table}
    \centering
    \caption{Pre-training code attack on the pre-trained models}
    \resizebox{7cm}{!}{%
    \begin{tabular}{ll|llll}
    \noalign{\smallskip}\toprule
    &                   &           \multicolumn{4}{c}{Memorization rate} \\
    \cmidrule{3-6}
    Model                   & Prefix length    & EM        & BLEU &METEOR &ROUGE-L    \\ 
    \cmidrule{1-6}
    \multirow{3}{*}{Starcoder2-3B} & 100 tokens & 0.389 & 0.652 & 0.706 & 0.682 \\
                                    & 150 tokens & 0.432 & 0.688 & 0.735 & 0.712 \\
                                    & 200 tokens & 0.462 & 0.715 & 0.756 & 0.737 \\
                                    & 250 tokens & 0.483 & 0.728 & 0.770 & 0.751\\
    \cmidrule{1-2}
    \multirow{3}{*}{Starcoder2-7B} & 100 tokens & 0.395	 & 0.663 & 0.720 &  0.694 \\
                                    & 150 tokens & 0.452 & 0.701 & 0.756 & 0.733 \\
                                    & 200 tokens & 0.485 & 0.730 & 0.772 & 0.753 \\
                                    & 250 tokens & 0.498 & 0.742 & 0.781 & 0.763 \\
    \cmidrule{1-2}
    \multirow{3}{*}{Starcoder2-15B} & 100 tokens & 0.455 & 0.714 & 0.760 & 0.738 \\
                                    & 150 tokens & 0.506 & 0.757 & 0.770 & 0.774 \\
                                    & 200 tokens & 0.536 & 0.777 & 0.806 & 0.794 \\
                                    & 250 tokens & 0.549 & 0.790 & 0.819 & 0.805 \\
    \noalign{\smallskip}\bottomrule
    \end{tabular}}
    \label{tab:attack-pre-forg-plen}
\end{table}

We observe that model size impacts pre-training extractability in pre-trained models.
Table~\ref{tab:attack-pre-forg-plen} indicates that using a 100-token prefix, we are able to extract 38.9\% of the pre-training code samples from the StarCoder2-3B model. 
This extraction rate increases to 45.5\% when using the largest version of the model, StarCoder2-15B.

\begin{RQanswer}
     \textbf{RQ1:} 
     The base version of StarCoder2 memorizes pre-training data, with memorization rate scaling with model size. 
     The extractability of this data increases with longer input prefixes, though not linearly.
\end{RQanswer}

\subsection{Extracting Pre-training Data From Fine-tuned Models}
\begin{figure}
    \centering
    \includegraphics[width=0.7\linewidth]{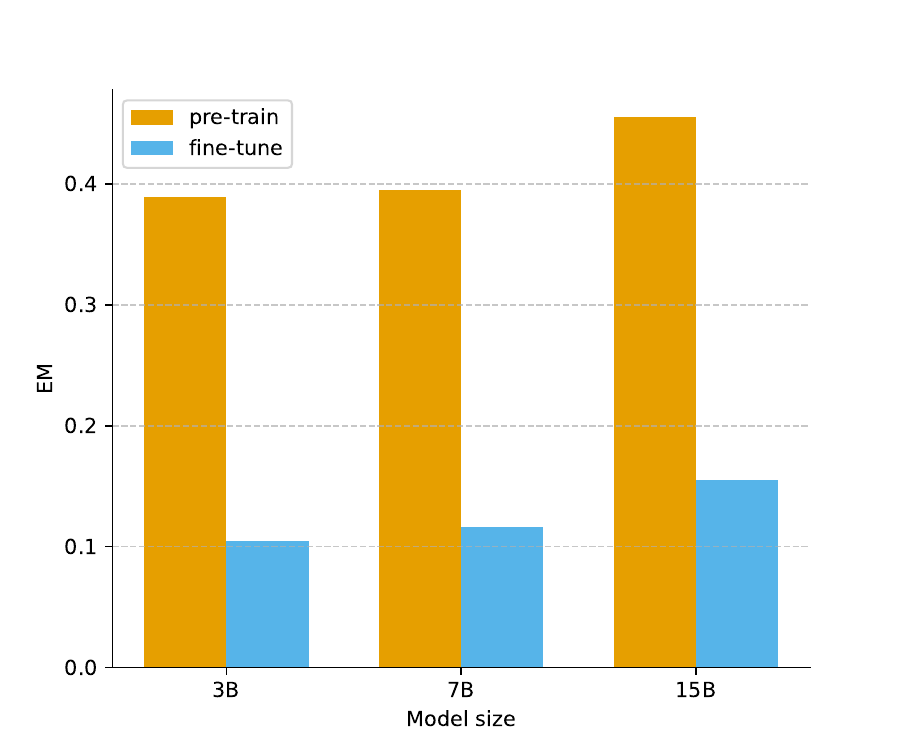}
    \caption{EM rate when performing the pre-training code attack (with 100 token prefix) on base and fine-tuned models.}
    \label{fig:forg_msize-pre-after}
\end{figure}

In Figure~\ref{fig:forg_msize-pre-after}, 
we plot the EM rates for the pre-training code attack on base and fine-tuned models.
We notice a considerable negative impact of fine-tuning on the extractability of pre-training data. 
Primarily, for the StarCoder2-15B model, we observe that after three epochs of fine-tuning, $30\%$ of the samples that were previously extractable during the pre-training stage could no longer be extracted. 
This trend is consistent across different model sizes to the same extent. 

Table~\ref{tab:attack-forge-ep} presents the results of experiments involving the pre-training code attack on fine-tuned models across the fine-tuning epochs, with the input prefix length fixed at 100 tokens.
The results show that fine-tuning epochs does not significantly decrease extractability. 
For the StarCoder2-15B base model, we initially extract 45.5\% of the pre-training code attack dataset. 
This rate decreases to 14.5\% after a single epoch of fine-tuning, with subsequent epochs maintaining a similar EM rate, showing only minor fluctuations within the standard error.
For the StarCoder2-3B and 7B models, the extractability of pre-training data decreases after the first epoch but increases slightly in subsequent epochs. 
Specifically, the EM rate for the StarCoder2-3B model drops to 8.7\% after the first fine-tuning epoch, increases to 10.8\% by the second epoch, and stabilizes at 10.5\% after the third epoch. 
The StarCoder2-7B exhibits a more pronounced initial decline, with the EM rate decreasing to 1.5\% after the first epoch, and then increasing to 11.0\% and 11.6\% after the second and third epochs, respectively.

\begin{table}
    \centering
    \caption{Pre-training code attack on fine-tuned models across several fine-tuning epochs.}
    \resizebox{7cm}{!}{%
    \begin{tabular}{ll|llll}
    \noalign{\smallskip}\toprule
    &                   &           \multicolumn{4}{c}{Memorization rate} \\
    \cmidrule{3-6}
    Model                   & Epochs    & EM        & BLEU   &METEOR &ROUGE-L\\ 
    \cmidrule{1-6}
    \multirow{3}{*}{Starcoder2-3B}  & 0 & 0.389 & 0.652 & 0.706 & 0.682 \\
                                    & 1 & 0.087 & 0.448 & 0.551 & 0.517 \\
                                    & 2 & 0.108 & 0.452 & 0.552 & 0.518 \\
                                    & 3 & 0.105 & 0.447 & 0.551 & 0.514 \\
    \cmidrule{1-2}
    \multirow{3}{*}{Starcoder2-7B} & 0 & 0.395 & 0.663 & 0.712 & 0.694 \\
                                    & 1 & 0.015 & 0.381 & 0.524 & 0.483 \\
                                    & 2 & 0.110 & 0.460 & 0.566 & 0.527 \\
                                    & 3 & 0.116 & 0.470 & 0.572 & 0.533 \\
    \cmidrule{1-2}
    \multirow{3}{*}{Starcoder2-15B} & 0 & 0.455 & 0.714 & 0.760 & 0.738 \\
                                    & 1 & 0.145 & 0.525 & 0.613 & 0.583 \\
                                    & 2 & 0.143 & 0.530 & 0.619 & 0.585 \\
                                    & 3 & 0.155 & 0.544 & 0.629 & 0.595 \\
    \noalign{\smallskip}\bottomrule
    \end{tabular}}
    \label{tab:attack-forge-ep}
\end{table}

\begin{table}
    \centering
    \caption{Pre-training code attack on fine-tuned models varying the input prefix length.}
    \resizebox{7cm}{!}{%
    \begin{tabular}{ll|llll}
    \noalign{\smallskip}\toprule
    &                   &           \multicolumn{4}{c}{Memorization rate} \\
    \cmidrule{3-6}
    Model                   & Prefix length    & EM        & BLEU &METEOR &ROUGE-L     \\ 
    \cmidrule{1-6}
    \multirow{3}{*}{Starcoder2-3B} & 100 tokens & 0.105 & 0.447  & 0.551 & 0.514 \\
                                    & 150 tokens & 0.124 & 0.503 & 0.598 & 0.566 \\
                                    & 200 tokens & 0.151 & 0.526 & 0.613 & 0.583 \\
                                    & 250 tokens & 0.150 & 0.534 & 0.618 & 0.592 \\
    \cmidrule{1-2}
    \multirow{3}{*}{Starcoder2-7B} & 100 tokens & 0.116	 & 0.470 & 0.572 & 0.533 \\
                                    & 150 tokens & 0.150 & 0.524 & 0.615 & 0.585 \\
                                    & 200 tokens & 0.171 & 0.546 & 0.631 & 0.600 \\
                                    & 250 tokens & 0.178 & 0.560 & 0.642 & 0.612 \\
    \cmidrule{1-2}
    \multirow{3}{*}{Starcoder2-15B} & 100 tokens & 0.155 & 0.544 & 0.629 & 0.595 \\
                                    & 150 tokens & 0.198 & 0.577 & 0.657 & 0.624 \\
                                    & 200 tokens & 0.220 & 0.600 & 0.675 & 0.645 \\
                                    & 250 tokens & 0.235 & 0.618 & 0.687 & 0.660 \\
    \noalign{\smallskip}\bottomrule
    \end{tabular}}
    \label{tab:attack-forge-plen}
\end{table}

Table \ref{tab:attack-forge-plen} shows the results of the experiments involving the pre-training code attack on the fine-tuned models while varying the input prefix length of the attacks. 
In line with the findings of the first RQ, when attacking the pre-trained models, extending the input prefix allows us to extract more pre-training samples from the fine-tuned models. 
For instance, with a 100-token input attack on StarCoder2-15B, we extract 15.5\% of the pre-training code attack samples. 
This extraction rate increases to 23.5\% when extending the input prefix to 250 tokens. 
Comparing the pre-training code attack on pre-trained models (Table~\ref{tab:attack-pre-forg-plen}) with that on fine-tuned models (Table~\ref{tab:attack-forge-plen}), we observe that the increase in extractability from extending the input prefix is less pronounced in fine-tuned models compared to pre-trained models. 
For instance, extending the input prefix from 100 to 150 tokens allowed the extraction of 43 additional samples from the StarCoder2-3B base model. 
In contrast, the same prefix length extension on the fine-tuned version resulted in the extraction of only 19 additional samples.

\begin{RQanswer}
     \textbf{RQ2:}
     Fine-tuning reduces the extractability of pre-training data.
     This effect is consistent across model sizes and unaffected by the number of training epochs. 
     Extending the input prefix length increases extractability but with a lower magnitude compared to pre-trained models.
\end{RQanswer}

\begin{table}
    \centering
    \caption{Fine-tuning code attack on fine-tuned models across fine-tuning epochs}
    \resizebox{7cm}{!}{%
    \begin{tabular}{ll|llll}
    \noalign{\smallskip}\toprule
    &                   &           \multicolumn{4}{c}{Memorization rate} \\
    \cmidrule{3-6}
    Model                   & Epochs    & EM        & BLEU   & EM $\uparrow$\% & BLEU $\uparrow$\% \\ 
    \cmidrule{1-6}
    \multirow{3}{*}{Starcoder2-3B} & 0 & 0.263 & 0.538 &  &  \\
                                    & 1 & 0.459 & 0.739 & 0.196 & 0.168 \\
                                    & 2 & 0.595 & 0.757 & 0.332 & 0.218 \\
                                    & 3 & 0.631 & 0.764 & 0.368 & 0.226\\
    \cmidrule{1-2}
    \multirow{3}{*}{Starcoder2-7B} & 0 & 0.249 & 0.546 &  &  \\
                                    & 1 & 0.408 & 0.739 & 0.159 & 0.158 \\
                                    & 2 & 0.527 & 0.741 & 0.278 & 0.194 \\
                                    & 3 & 0.578 & 0.749 & 0.329 & 0.202 \\
    \cmidrule{1-2}
    \multirow{3}{*}{Starcoder2-15B} & 0 & 0.300 & 0.576 &  &   \\
                                    & 1 & 0.549 & 0.756 & 0.249 & 0.181  \\
                                    & 2 & 0.567 & 0.761 & 0.267 & 0.185 \\
                                    & 3 & 0.609 & 0.774 & 0.309 &  0.198 \\
    \noalign{\smallskip}\bottomrule
    \end{tabular}}
    \label{tab:attack-mem-ep}
\end{table}

\begin{figure}
    \centering
    \includegraphics[width=0.7\linewidth]{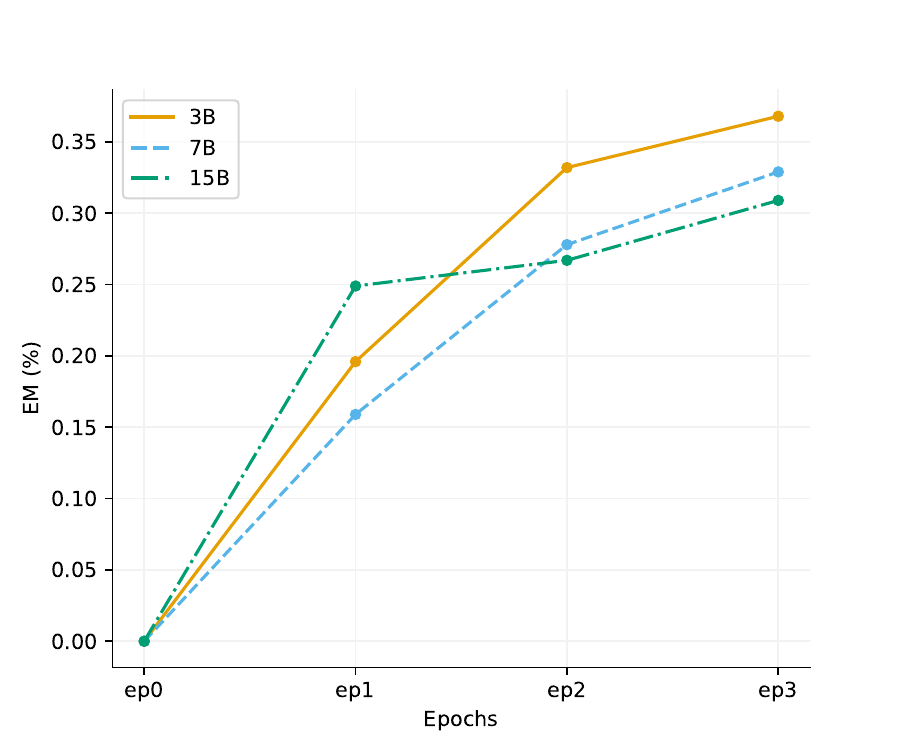}
    \caption{EM rate when performing the fine-tuning code attack (with 100 token prefix) across fine-tuning epochs.}
    \label{fig:mem_ep}
\end{figure}

\subsection{Extracting Fine-tuning Data}
In Table \ref{tab:attack-mem-ep}, we present the results of experiments involving the fine-tuning code attack across the fine-tuning epochs. 
Notably, performing the fine-tuning code attack on the pre-trained models revealed that some fine-tuning samples were also retrievable from the pre-trained models. 
In this section, we demonstrate the increase in EM and BLEU ($\uparrow$\%) relative to the baseline at epoch zero. Due to space limits, METEOR and ROUGE-L scores are available in the replication package~\cite{rep_package}.
The baseline represents the EM rate achieved by executing the fine-tuning code attack on the pre-trained models. 
Our experiments reveal that extractability increases with the number of fine-tuning epochs. 
Specifically, after one epoch of training StarCoder2-3B, we were able to extract 45.9\% of the fine-tuning samples.
This extraction rate increased to 59.5\% after the second epoch and further to 63.1\% after the third epoch. 

In Figure~\ref{fig:mem_ep}, we present the increase of the EM rate over the baseline across epochs.
Our experiments reveal that smaller models are more prone to extraction on fine-tuning data than bigger ones.
Based on Figure~\ref{fig:mem_ep}, we can see that as the number of fine-tuning epochs increases, smaller models exhibit greater vulnerability to data extractability, whereas larger models demonstrate reduced exposure. 
Specifically, when attacking the StarCoder2-3B model after three epochs of fine-tuning, we extracted 36.8\% additional samples relative to the baseline. 
In contrast, attacking the StarCoder2-15B model under the same conditions yielded 30.9\% additional extractable samples compared to the baseline.

Table~\ref{tab:attack-mem-dedup} shows the results of the experiments consisting of attacking the fine-tuned models controlling the duplication rate (i.e., the frequency of 300-token samples in the fine-tuning corpus). 
We find that the duplication rate has a notable impact on extractability. 
For StarCoder2-3B, extractability is below 5\% with unique samples. 
This percentage increases to 23.7\% with a single duplicate in the fine-tuning corpus and further to 45.1\% with two duplicates. 

\begin{table}
    \centering
    \caption{Fine-tuning code attack on fine-tuned models with varying duplication rates.}
    \resizebox{7cm}{!}{%
    \begin{tabular}{ll|llll}
    \noalign{\smallskip}\toprule
    &                   &           \multicolumn{4}{c}{Memorization rate} \\
    \cmidrule{3-6}
    Model                   & Duplication    & EM        & BLEU &EM$\uparrow$\% &BLEU$\uparrow$\%  \\ 
    \cmidrule{1-6}
    \multirow{3}{*}{Starcoder2-3B} & =1 & 0.049 & 0.423 & 0.025 & 0.074 \\
                                    & =2 & 0.237 & 0.615 & 0.172 & 0.174\\
                                    & =3 & 0.451 & 0.754 & 0.329 & 0.333 \\
                                    & $>$3 & 0.631 & 0.764 & 0.368 & 0.266 \\
    \cmidrule{1-2}
    \multirow{3}{*}{Starcoder2-7B} & =1 & 0.061 & 0.433 & 0.042 & 0.082\\
                                    & =2 & 0.238 & 0.616 & 0.181 & 0.174 \\
                                    & =3 & 0.450 & 0.776 & 0.348 & 0.337\\
                                    & $>$3 & 0.578 & 0.749 & 0.329 & 0.202\\
    \cmidrule{1-2}
    \multirow{3}{*}{Starcoder2-15B} & =1 & 0.075 & 0.475 &0.052 &0.110 \\
                                    & =2 & 0.352 & 0.694 &0.277 & 0.244 \\
                                    & =3 & 0.544 & 0.832 &0.417 & 0.385 \\
                                    & $>$3 & 0.609 & 0.774 &0.309 & 0.198  \\
    \noalign{\smallskip}\bottomrule
    \end{tabular}}
    \label{tab:attack-mem-dedup}
\end{table}

Table~\ref{tab:attack-mem-plen} illustrates the effect of extending the input prefix length on the success of fine-tuned code extraction attacks. 
 Our results indicate that the extractability of fine-tuning data increases with the input prefix length.
We can extract 63.1\% of the samples from StarCoder2-3B inputting 100 prefix tokens. 
This percentage increased to 68.9\% with a prefix of 150 tokens and further to 81.5\% with a prefix of 250 tokens.

\begin{table}
    \centering
    \caption{Fine-tuning code attack on fine-tuned models with varying prefix lengths.}
    \resizebox{7cm}{!}{%
    \begin{tabular}{ll|llll}
    \noalign{\smallskip}\toprule
    &                   &           \multicolumn{4}{c}{Memorization rate} \\
    \cmidrule{3-6}
    Model                   & Prefix length    & EM        & BLEU &EM$\uparrow$\% &BLEU$\uparrow$\%    \\ 
    \cmidrule{1-6}
    \multirow{3}{*}{Starcoder2-3B} & 100 tokens & 0.631 & 0.764 & 0.367 & 0.226 \\
                                    & 150 tokens & 0.689 & 0.785 & 0.393 & 0.231 \\
                                    & 200 tokens & 0.747 & 0.816 & 0.443 & 0.260 \\
                                    & 250 tokens & 0.815 & 0.865 & 0.502 & 0.297\\
    \cmidrule{1-2}
    \multirow{3}{*}{Starcoder2-7B} & 100 tokens & 0.578	 & 0.750 & 0.330 & 0.202 \\
                                    & 150 tokens & 0.646 & 0.773 & 0.386 & 0.220 \\
                                    & 200 tokens & 0.712 & 0.802 & 0.436 & 0.243 \\
                                    & 250 tokens & 0.800 & 0.850 & 0.492 & 0.245 \\
    \cmidrule{1-2}
    \multirow{3}{*}{Starcoder2-15B} & 100 tokens & 0.610 & 0.775 & 0.310 & 0.198 \\
                                    & 150 tokens & 0.730 & 0.805 & 0.372 & 0.216 \\
                                    & 200 tokens & 0.817 & 0.830 & 0.441 & 0.232 \\
                                    & 250 tokens & 0.888 & 0.868 & 0.474 & 0.244 \\
    \noalign{\smallskip}\bottomrule
    \end{tabular}}
    \label{tab:attack-mem-plen}
\end{table}

\begin{RQanswer}
     \textbf{RQ3:}
    Smaller models are more prone to extraction of fine-tuning data. Duplication rate significantly affects extractability. Increasing fine-tuning epochs leads to more samples memorized, and extending the input prefix length also facilitates fine-tuning extractability.
\end{RQanswer}

\subsection{Type of Extracted Data}
Table~\ref{tab:attack-tags} represents the type of code data extracted through our attacks. 
During the tagging process, we find numerous examples of names, emails, and GitHub links memorized by the model~\cite{rep_package}.

\textbf{\begin{table}
    \centering
    \caption{Code attack results per data category and model.}
    \resizebox{7cm}{!}{%
    \begin{tabular}{ll|ll|l}
    \noalign{\smallskip}\toprule
    \multicolumn{2}{c}{\textbf{Code Attack type}} &  \multicolumn{2}{c}{Pre-training (RQ1-2)}  & Fine-tuning (RQ3)  \\
    \cmidrule{1-5}
    Model                   & Category    & Base       & Fine-tuned & Fine-tuned  \\ 
    \cmidrule{1-5}
    \multirow{4}{*}{Starcoder2-3B} & Code & 115    & 53  & 56  \\
                                    & Dicts & 124  & 31  & 54  \\
                                    & Docs & 32    & 4   & 104  \\
                                    & License & 85 & 8   & 185   \\
                                    & Test & 3     & 2   & 1   \\
    \cmidrule{1-2}
    \multirow{4}{*}{Starcoder2-7B} & Code  & 117   & 64 & 53  \\
                                    & Dicts & 112  & 26 & 52  \\
                                    & Docs  & 42   & 5  & 123  \\
                                    & License & 88 & 11 & 181  \\
                                    & Test    & 3  & 3  & 1  \\
    \cmidrule{1-2}
    \multirow{4}{*}{Starcoder2-15B} & Code & 136   & 62 & 88 \\
                                    & Dicts & 133  & 47 & 64  \\
                                    & Docs & 52    & 16 & 137  \\
                                    & License & 94 & 15 & 195  \\
                                    & Test & 4     & 4  & 1  \\
    \noalign{\smallskip}\bottomrule
    \end{tabular}}
    \label{tab:attack-tags}
\end{table}}

\begin{figure}
    \centering
    \includegraphics[width=0.7\linewidth]{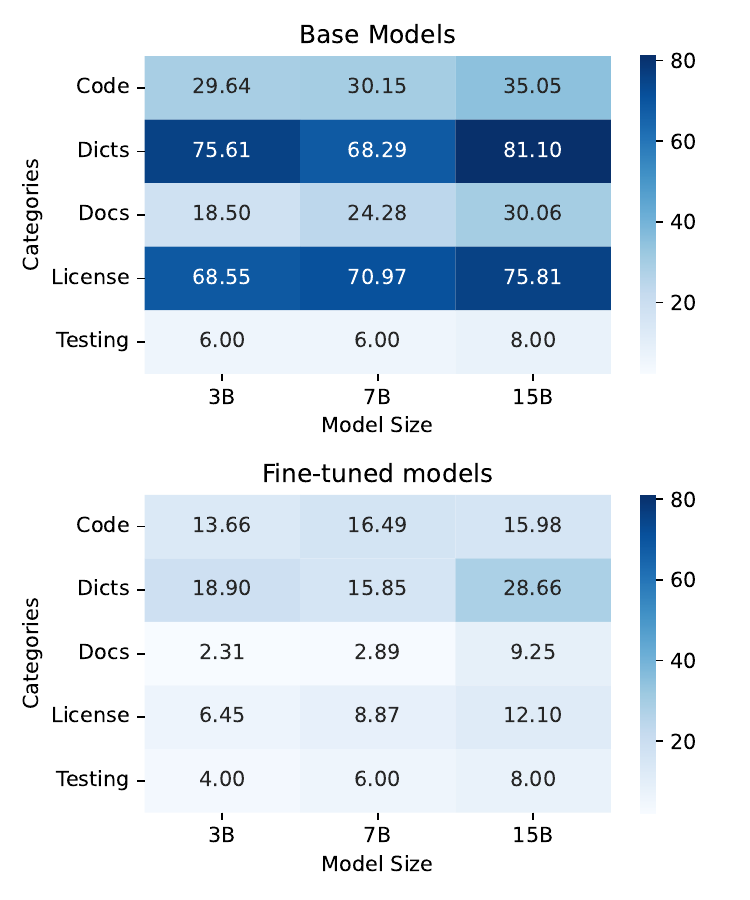}
    \caption{Percentage of extractable pre-training samples per category on base and fine-tuned models.}
    \label{fig:tag_pretr}
\end{figure}

Figure~\ref{fig:tag_pretr} shows the percentage per category of memorized samples performing the pre-training code attack on the base and fine-tuned models, respectively. 
The results indicate that \textit{data carriers} and \textit{licenses} are the most frequently memorized pre-training samples. 
Fine-tuning has an impact on the extractability of pre-training data. 
We find that this rate is not uniform between categories. 
Some categories are less extractable than others. 
Specifically, licenses become less extractable after fine-tuning compared to data carriers. 
For instance, the StarCoder2-3B base model memorizes 75.61\% of the data carriers in the pre-training code attack dataset; this percentage decreases to 18.90\% following fine-tuning. 
Similarly, the base model retains 68.55\% of the licenses, 
but this figure drops to 6.45\% after fine-tuning. 
Data carriers exhibit the highest memorization rate post-fine-tuning, being the category of code less affected by the ``forgetting'' phenomenon observed when performing the pre-training code attack on fine-tuned models.

The last column of Table~\ref{tab:attack-tags} 
presents the amount of memorized samples per category from the fine-tuning code attack. 
The results indicate that similar to pre-training code attacks on pre-trained models, data carriers and licenses are the most memorized samples after fine-tuning. 
Specifically, the fine-tuned StarCoder2-15B model retained 54.70\% of the data carriers and 68.66\% of licenses from the fine-tuning code attack dataset.

\begin{RQanswer}
     \textbf{RQ4:} Data carriers and license information are the most likely to be memorized, while the latter is the most likely to be forgotten after fine-tuning.   
\end{RQanswer}

\section{Discussion}
\subsection{Pre-training Data Memorization - Pre-trained Models}
Our findings indicate that the StarCoder2 base model family tends to memorize its pre-training data. 
Consistent with prior research~\cite{carlini2022quantifying, alkaswan2024traces, yang2024unveiling}, we observed that the extent of memorization correlates with model size. 
Using a 100-token attack on the StarCoder2-15B model, we extracted 45.5\% of the dataset. This rate dropped to 38.9\% with the three-billion-parameter version.
Our results substantiate the log-linear relation between size and memorization, which has been recently observed also for LLMs trained on code~\cite{alkaswan2024traces}.

Our experiments reveal that extending the input prefix length enhances the memorization rate of the pre-trained models when attacking with samples from the pre-training data.
These findings align with previous studies, which demonstrated that longer input prefixes facilitate the retrieval of more information from the model~\cite{carlini2022quantifying, yang2024unveiling}. 
Our results reinforce the connection identified by \citeauthor{carlini2021extracting}~\cite{carlini2021extracting} between extractability and the precise identification of key information necessary for accessing specific details from a language model~\cite{carlini2021extracting}. 

\citeauthor{alkaswan2024traces}~\cite{alkaswan2024traces} demonstrated that data carriers are memorized at a higher rate compared to regular code or documentation~\cite{alkaswan2024traces}. 
Our findings are consistent with this observation, as our categorization revealed that the types of code most frequently memorized by the pre-trained models were data carriers and licenses. 
We extracted up to 80\% and 75\% of the total data carriers and licenses from the attack dataset, but only 30\% for code and documentation.

\subsection{Pre-training Data Memorization - Fine-tuned Models}
Performing the pre-training code attack on the fine-tuned models revealed that fine-tuning reduces the extractability of pre-training data.
We observed that after three epochs of fine-tuning, around 30\% of the samples that were previously extracted during the pre-training stage could no longer be extracted. 
To further investigate this phenomenon, we replicated the experiments conducted on the pre-trained models to examine how factors influencing pre-training extractability in pre-trained models affect fine-tuned models.

Our experiments revealed that the decrease in extractability 
is consistent across different model sizes to the same extent. 
Additional fine-tuning epochs do not significantly decrease extractability; despite some fluctuations, the EM rate does not consistently drop across epochs. 
For the 3B and 7B parameter models, extractability after the first epoch is slightly lower compared to later epochs. 
For example, the EM rate for the StarCoder2-3B model is 8.7\% after the first fine-tuning epoch, increasing to 10\% by the second epoch and remaining stable through the third epoch. 
This pattern likely reflects the model's initial adjustment to new, unseen data during the first epoch of fine-tuning~\cite{NEURIPS2022_0cde695b}. 
The model requires a few epochs to adapt to this new data distribution and perform the code completion task effectively.
The initial drop at the first epoch is more pronounced in the 7B model.
\citeauthor{lozhkov2024starcoder2stackv2} state that is not clear why StarCoder2-7B is not performing as well as the 3B and 15B models for their size~\cite{lozhkov2024starcoder2stackv2}. 
We considered this as a plausible explanation for the observed discrepancies.
Note that after the initial drop, extractability stabilizes in subsequent epochs.

Inspecting the type of pre-training code that is no longer extractable by fine-tuned models, 
showed that some categories are less extractable than others. 
Specifically, licenses become less extractable after fine-tuning compared to data carriers. 
StarCoder2-3B base model memorized 68.55\% of the licenses, but this figure dropped to 6.45\% after fine-tuning. 
Notably, data carriers exhibit the highest memorization rate post-fine-tuning, suggesting that this category of code is less affected by the ``forgetting'' phenomenon observed in the fine-tuned models.

\subsection{Fine-tuning Data Memorization}
Performing the fine-tuning code attack revealed that, even only one training epoch is sufficient to make the model memorize 55\% of the attack dataset.
This number goes up to 61\% after the third epoch.
Overall, extractability increases at each fine-tuning epoch.

Attacking different model sizes, allowed us to observe that smaller models memorize more fine-tuning data.
As shown in Figure \ref{fig:mem_ep}, this behavior happened only from the second fine-tuning epoch.
Attacking the fine-tuned three billion model we were able to extract 36.8\% of additional samples. 
This number decreased to 30.9\% attacking the fine-tuned fifteen billion model.

\citeauthor{kandpal2022deduplicating, carlini2022quantifying} revealed that the duplication rate severely impacts extractability.
This result also persists in fine-tuning memorization.
For StarCoder2-3B, extractability is below 5\% with unique samples (i.e., $d=1$). 
This percentage rises to 23.7\% with a single duplicate in the fine-tuning corpus and further to 45.1\% with two duplicates (i.e., $d=3$). 

Performing the fine-tuning code attack on the pre-trained models revealed that some fine-tuning samples were also retrievable from the pre-trained models. 
Several factors may contribute to this observation. 

Notably, it indicates that even before fine-tuning, the pre-trained models had already memorized portions of the fine-tuning dataset, despite the extensive filtering process applied to the fine-tuning data~\cite{katzy2025heapcontaminationfreemultilingualcode}.
The repetitive and highly structured nature of Java syntax, which adheres to specific patterns that enhance readability, maintainability, and consistency, could be a significant factor. 
Additionally, our approach to constructing the attack dataset did not account for code complexity, leading to some suffixes being easily inferred by the model.

To further investigate, we inspected the types of code extractable from the pre-trained models. 
We found that the majority of the fine-tuning samples retrievable from the pre-trained models consisted of licenses. Specifically, performing the fine-tuning code attack on the StarCoder2-3B base model, we extracted 120 samples out of 917, of which 107 were licenses.

\subsection{Implications}
\textit{Risks for Fine-tuning Models}:
The study's findings emphasize key implications for fine-tuning LLMs on code data. Firstly, models' tendency to memorize training data poses significant risks, especially with proprietary or sensitive datasets. Fine-tuning on smaller, task-specific datasets increases exposure to memorization and data leakage.
Smaller models are more prone to memorizing fine-tuning data compared to larger models. Although they are cheaper and faster to fine-tune, they carry a higher risk of data leakage. Therefore, companies and individuals fine-tuning code language models on personal or proprietary data should consider this risk when selecting the model size.

\textit{Security Vulnerability Propagation}:
LLMs may generate code snippets containing security flaws present in their training data, potentially introducing vulnerabilities into new software projects. This could lead to the unintentional spread of security issues across codebases~\cite{alkaswan2023abuse, alkaswan2024traces, yang2024unveiling}.

\textit{Data Privacy Leakage Concern}:
If the model outputs verbatim snippets from proprietary or sensitive data used in fine-tuning, it could lead to unintended disclosure of confidential information. This is particularly problematic for industries dealing with personal, medical, or financial data. This also raises ethical questions about consent, especially if the data involves personal information or creative works used without explicit permission for such reproduction~\cite{henderson2023foundation}. 

\textit{Legal and IP Issues}:
Many existing LLMs for code utilize copyleft and other non-permissive licensed code~\cite{alkaswan2023abuse}. Using public code to train these models can fall under fair use, which allows copyrighted works to be used in new and transformative ways in many jurisdictions~\cite{henderson2023foundation}. However, if the model's output closely resembles the copyrighted input, fair use may no longer apply. In such cases, the output must adhere to the original license terms, which may include share-alike and attribution requirements~\cite{henderson2023foundation}.
Outputting memorized samples from proprietary datasets could potentially infringe on intellectual property rights. This raises significant legal concerns, especially if the model is used commercially or made publicly accessible. Ensuring compliance with IP laws is crucial to avoid litigation and uphold ethical standards in model deployment~\cite{choksi2023textanywayexploringbigcode}.  

\textit{Overfitting and Reduced Generalization}:
If the model memorizes specific examples from the fine-tuning data instead of learning general patterns, it may perform poorly on new, unseen data. This overfitting can limit the model's effectiveness in real-world applications~\cite{hartmann2023sokmemorizationgeneralpurposelarge, carlini2021extracting}.

\subsection{Recommendations}
To mitigate these risks, organizations that fine-tune foundation models on specialized or proprietary data need to implement robust safeguards.
From a technical perspective, implementing differential privacy techniques like Whispered Tuning~\cite{singh2024whispered} can help ensure that the model does not memorize specific training data by adding noise during the training process. Additionally, using data filtering and deduplication methods can remove sensitive information and prevent the inclusion of duplicate data entries, thereby reducing the risk of memorization~\cite{kandpal2022deduplicating, lee2022deduplicating, carlini2022quantifying}. When fine-tuning models, it is important to limit the number of epochs, employ regularization techniques, and prefer larger models to prevent overfitting and memorization.

Organizational practices are also vital in ensuring the secure and ethical deployment of LLMs. Active monitoring of model outputs through automated checks is essential to detect and address any instances where the model might reproduce memorized data. Developing methods to ``forget'' specific pieces of information post-training, such as unlearning techniques, can further safeguard against data leakage~\cite{alkaswan2024traces, yang2024unveiling}.

Memorization can put creators and users of LLMs for code at legal risk, including license and fair use infringements~\cite{alkaswan2023abuse, henderson2023foundation}. To mitigate these risks, it is crucial to train models using code licensed under permissive licenses (such as BSD-3 or MIT) or to provide provenance information that traces the code back to its source~\cite{alkaswan2023abuse, alkaswan2024traces, yang2024unveiling, henderson2023foundation}.

\subsection{Ethical considerations}
Although this work explores techniques that could potentially be used to extract sensitive information from models, it is conducted with ethical considerations in mind. 
Our objective is to highlight the issue of memorization in LLMs for code, inform users and creators about this problem, and provide tools for measuring it. 
We emphasize that our work does not intentionally expose private information and we strongly urge users of our work to avoid such practices.
We believe the benefits of sharing our data outweigh the risks, and we decide to make them available.

\subsection{Limitations and Threats to the Validity}
\subsubsection{Internal Validity}
In our evaluation, we did not account for the spatial distribution of samples within the files. 
While the samples maintain a fixed token length, their locations within the files are selected arbitrarily to avoid bias. 
The methodology was chosen to ensure samples could be extracted from any location within the file
to reflect a realistic and varied scenario~\cite{alkaswan2024traces, carlini2022quantifying}.

\subsubsection{External Validity}
While we study three model sizes, 
our evaluation focuses on the StarCoder2 model family. 
Other model architectures may exhibit different levels of memorization.
Additionally, our pre-training code dataset considers samples that are duplicated more than three times in the training corpus. 
Due to computational constraints, we were not able to control the duplication rate as we did on the fine-tuning code attack dataset. 
More studies on pre-training code attacks on low-duplication data could extend our findings.

Most previous studies have centered on Python. This leaves a gap for other programming languages.
We focused on Java for our attack datasets due to its well-defined syntax and structure, which may impact memorization. Given Java's popularity, our results may not apply to less common languages. In future evaluations, we plan to include more programming languages to compare how different languages are memorized by the model.

\subsubsection{Construct Validity}
We primarily use the EM metric for reporting extractability rates in code LMs as we focus on exact reproductions to address privacy and security concerns associated with memorization. However, this metric likely underestimates the true extent of memorized samples, as some might be slightly altered by the model~\cite{hartmann2023sokmemorizationgeneralpurposelarge}. 
To mitigate this threat, 
we also compute and report other fuzzy metrics such as BLEU, METEOR, and ROUGE-L in the paper and in our replication package~\cite{rep_package}. Our findings indicate that these metrics are highly correlated with the EM rate. 

\section{Conclusion}
In this study, we conducted an extensive investigation into data extraction from pre-trained and fine-tuned code language models.
We formally defined a data extraction security game based on the concepts of k-extractability and membership inference attacks. 
This framework was employed to develop a custom benchmark for assessing the vulnerability of both pre-training and fine-tuning samples to extraction attacks.
Our findings indicate that the StarCoder2 model memorizes its pre-training data. 
We further observed that fine-tuning reduces the extractability of pre-training data. 
Contrarily, we found that fine-tuning increases the vulnerability of smaller models to fine-tuning data extraction attacks compared to larger models. 
Additionally, our results show that data carriers and license information are the most likely to be memorized, with the latter being the most likely to be forgotten after fine-tuning.
These insights emphasize the importance of addressing memorization during fine-tuning to safeguard against potential data leaks and highlight the need for further research and development of strategies to mitigate these vulnerabilities.

\printbibliography
\end{document}